\documentclass[final,3p,times]{elsarticle}

\usepackage{amssymb}
\usepackage{amsmath}
\usepackage{multirow}
\usepackage{caption}
\usepackage{subcaption}

\journal{ARXIV}

\begin{document}

\begin{frontmatter}



\title{Estimating the mass of the thin shell of gravastars in generalised cylindrically symmetric space-time within the framework of Rastall theory of gravity}


\author{Debadri Bhattacharjee \& Pradip Kumar Chattopadhyay} 

\address{IUCAA Centre for Astronomy Research and Development (ICARD), Department of Physics, Cooch Behar Panchanan Barma University, Vivekananda Street, District: Cooch Behar,  Pin: 736101, West Bengal, India.}
\ead{debadriwork@gmail.com \& pkc$_{-}$76@rediffmail.com}

\begin{abstract}
This study investigates the gravastars in the framework of Rastall theory of gravity in generalised cylindrically symmetric space-time. Following the Mazur-Mottola hypothesis (P. O. Mazur and E. Mottola, Universe {\bf 9}, 88 (2023)), gravastars are classified as one of the most unique and exotic kind of compact objects, presenting themselves as a plausible alternative to black holes. In this study, we build upon the Mazur-Mottola framework of Gravitational Bose-Einstein Condensate (GBEC) stars by generalising it to a cylindrically symmetric spacetime within the framework of Rastall gravity to present a novel approach for estimating the mass limit of the thin shell of isotropic gravastars. We have ensured singularity-free solutions for the interior de-Sitter core, non-vanishing solutions for the thin shell and flat vacuum solution of the exterior region, within this parameter space. Under the framework of Rastall gravity and cylindrically symmetric spacetime, the Lanczos equations at the hypersurface junction $(r=R)$ undergo significant modifications, leading to a revised form of the Darmois-Israel junction conditions. These modified junction conditions are utilised to investigate the influence of the Rastall parameter $(\xi)$ on the mass of the thin shell and key characteristics of gravastars, including the shell's proper length, energy, and entropy. Additionally, we propose a novel method for estimating the mass of the thin shell using the concept of surface redshift $(Z_{s})$. By adhering to the Buchdahl upper limit, $Z_{s}<2$ for isotropic configuration, we have determined the mass bounds of the thin shell for various characteristic radii and values of the Rastall parameter $(\xi)$.   
\end{abstract}




\end{frontmatter}

\section{Introduction}\label{sec1} 
Classical black hole (BH) solution derived by K. Schwarzschild \cite{Schwarzschild} originated from the vacuum solutions of Einstein field equations (EFE) \cite{Einstein}. There are two main persistent issues with such solution, {\it viz.}, the presence of singularities at the centre $(r\rightarrow0)$ and at the event horizon $(r=\frac{2GM}{c^{2}})$, where $M$ is the total mass, $G$ is the gravitational constant and $c$ is the velocity of light. Now the singularity problem at the event horizon is termed the kinematical singularity and it can be removed by distinct coordinate transformations, such as Tortois coordinate, Eddington-Finkelstein coordinates, Kruskal-Szekeres coordinates etc. However, the central singularity or the dynamical singularity is irremovable. Apart from these, from a quantum mechanical perspective, a photon having an asymptotic frequency acquires a diverging energy density near the event horizon and there are no a-priori parameters that can sustain the deformation of the local geometry in that vicinity. In the semi-classical approximation, quantising a massless field in a fixed Schwarzschild background results in BH radiation with a thermal spectrum characterised by the asymptotic Hawking temperature, $T_{H}=\frac{\hbar}{8\pi k_{B}GM}$ \cite{Hawking}, where $\hbar$ and $k_{B}$ are the Planck's constant and Boltzmann constant. Further, semi-classical approximation assumes that the black hole's geometry remains essentially classical while quantum fields propagate within it, i.e., the back-reaction of this BH radiation on the classical spacetime is assumed to be minimal. However, a detailed calculation of the energy-momentum tensor exhibits an infinite blue-shift at the horizon \cite{Birrell}. The characteristic wavelengths contributing to these quantum stresses are of the order of $\frac{2GM}{c^{2}}$, indicating a non-local behaviour on this scale. When such a diverging energy density and pressure are incorporated into the semi-classical Einstein equations, it becomes evident that the geometry near the horizon undergoes significant deviations from the classical Schwarzschild solution. Unlike the kinematic singularity, these non-local semi-classical effects near the horizon cannot be eliminated through local coordinate transformations. Moreover, in the expression of Hawking temperature, the inverse dependence of $T_{H}$ and $M$ shows that a BH being in thermal equilibrium with its own Hawking temperature has negative specific heat. Therefore, it is unstable under thermodynamic fluctuations. 

To effectively resolve these issues, alternative models for the end state of gravitational collapse were necessary. In this regard, Mazur and Mottola extended the notion of Gravitational Bose-Einstein Condensate (GBEC) to formulate a new class of compact objects, termed Gravitational Vacuum Star or Gravastar \cite{Mazur,Mazur1,Mazur2}. In their preliminary model, gravastar had five layers which was modified to a three layered configuration by Visser and Wiltshire \cite{Visser}. Gravastars are widely recognised to possess a three-layered structure, comprising an inner region, an outer region, and an intermediate thin shell that separates these two domains. Each region is described by a distinct equation of state (EoS). The interior region, characterised by a de-Sitter condensate phase, exhibits positive pressure and negative energy density $(p=-\rho)$, effectively representing a positive cosmological constant. The negative energy density implies that the interior exerts an outward repulsive pressure at every point on the thin shell. The thin shell itself is proposed to consist of an ultra-relativistic stiff fluid, governed by the EoS $p=\rho$, which aligns with the Zel'dovich criterion for an EoS in the context of a cold baryonic universe \cite{Zeldovich, Zeldovich1}. In this region, the sound speed matches the speed of light, describing the extreme end of causality. The shell has a slice of finite thickness. In contrast, the exterior region corresponds to a vacuum flat spacetime, described by the EoS, $p=\rho=0$. 

Mazur and Mottola in their proposed model \cite{Mazur,Mazur1,Mazur2} solved the EFE for the three layers and found non-singular and non-vanishing solutions for the interior and shell region respectively. Apart from that they studied the stability of such exotic form of compact objects through the maximisation of entropy. Based on their results, they manifested gravastars as a viable alternative to BH. This alternative definition to BH formalism is an well established approach now in the study of exotic compact objects. Mazur and Mottola successfully demonstrated the thermodynamic stability of their configuration, the issue of dynamical stability under spherically symmetric perturbations of matter or gravitational fields remained somewhat unresolved. Visser and Wiltshire \cite{Visser} in their study developed a model that retained the essential characteristics of the Mazur-Mottola framework while being sufficiently simplified to allow for a comprehensive dynamical analysis. Their findings revealed that certain physically plausible EoS for the transition layer can indeed result in stability. In the context of stability, Chirenti and Rezzolla \cite{Chirenti} addressed two fundamental questions in their study: whether a gravastar represents a stable configuration and how it can be distinguished from a BH. Carter \cite{Carter} developed a gravastar model featuring a generalised Schwarzschild-(anti)-de Sitter or Reissner-Nordstr\"om exterior spacetime. Ray et al. \cite{Ray} demonstrated that for a gravastar to be physically viable within the framework of General Relativity (GR), the presence of pressure anisotropy and adherence to specific energy conditions are essential. Horvat and Ilijic \cite{Horvat} derived the energy conditions for a gravastar and established a stringent constraint for maintaining the dominant energy condition at the thin shell. In another study, DeBenedictis et al. \cite{DeBenedictis} successfully constructed a stable gravastar model characterised by continuous pressure and density profiles. Celine et al. \cite{Cattoen} investigated gravastar structures with pressure anisotropy, avoiding the infinitesimal thin shell approximation. Employing a distinct approach, Ghosh et al. \cite{Ghosh} formulated gravastar models using the Kuchowich metric potential. Gravastar configurations have also been explored in higher-dimensional contexts. Rahaman et al. \cite{Rahaman} presented a gravastar model in (2+1) dimensions, while in other work, Rahaman et al. \cite{Rahaman1} extended the analysis to D-dimensional Einstein gravity. Additionally, Ghosh et al. \cite{Ghosh1} developed a gravastar model in (3+1) dimensions within the framework of the Karmakar class-I condition. Bhar \cite{Bhar} examined the formation of higher-dimensional charged gravastars under conformal motion. In an innovative approach, Bilic et al. \cite{Bilic} replaced the de Sitter interior with a Born-Infeld phantom field in their gravastar model, drawing inspiration from low-energy string theory. Motivated by the concept of cosmic expansion, Lobo \cite{Lobo} analysed a gravastar solution with the interior governed by a dark energy equation of state, enabling the study of various dark energy stellar parameters.  

Modified gravity theories offer a compelling framework for understanding the formation and evolution of cosmic structures. While Einstein's General Relativity (GR) remains a cornerstone for unravelling the universe's mysteries, observational evidence supporting an accelerating universe, dark energy, and dark matter presents significant challenges to its theoretical predictions \cite{Riess,Khlopov,Khlopov1,Khlopov2,Riess1,Perlmutter,Bernardis,Hannay,Peebles,Padmanabhan,Sahni,Clifotn,Amanullah,Komatsu,Tegmark}. The necessity of incorporating modified gravity theories has been thoroughly discussed in \cite{Nojiri}. Generalizing the Einstein-Hilbert action has paved the way for various alternative gravity frameworks, including $f(R)$ gravity (where $R$ is the Ricci scalar) \cite{Capozziello,Elizalde}, $f(\mathcal{T})$ gravity (with $\mathcal{T}$ representing the torsion scalar) \cite{Bamba}, $f(R,\Box R,T)$ gravity (where $\Box$ denotes the D'Alembert operator and $T$ the trace of the energy-momentum tensor) \cite{Houndjo,Yousaf1,Ilyas}, and $f(R,T)$ gravity \cite{Harko}, among others. The gravastar paradigm has been extensively explored within the context of these modified gravity theories. For instance, Das et al. \cite{Das} investigated gravastar structures within the framework of $f(R,T)$ gravity. Bhar and Rej \cite{Bhar1} analysed stable and self-consistent charged gravastars in $f(R,T)$ gravity using conformal killing vectors. Additionally, Ghosh et al. \cite{Ghosh2} constructed gravastar solutions under the $f(T,T)$ gravity framework. In the context of $f(T)$ gravity, Das et al. \cite{Das1} investigated the structure of gravastars. Adopting a distinct approach, Sengupta et al. \cite{Sengupta} analysed gravastars within the framework of braneworld gravity, while Banerjee et al. \cite{Banerjee} utilised Finslerian geometry to examine gravastar configurations. Notably, Bhattacharjee et al. \cite{Bhattacharjee} developed a gravastar model under cylindrical symmetry and determined the thin shell mass within stability constraints. Furthermore, Shamir and Zia \cite{Shamir} explored the evolution of mass and radius for gravastars using $f(R,G)$ gravity. Additionally, various studies have advanced the understanding of compact astrophysical objects through the application of different modified gravity theories \cite{Ditta,Shamir1,Malik,Shamir2,Ahmed,Shamir3}. 

Unlike GR, one of the notable aspects of these alternative theories is the non-conservation of the energy-momentum tensor, which results in a non-minimal coupling between geometry and matter fields. This distinctive feature of a non-vanishing covariant divergence inspired the development of a charged gravastar model within the framework of Rastall gravity, originally proposed by P. Rastall \cite{Rastall,Rastall1}. In Rastall theory, the covariant derivative of the energy-momentum tensor is proportional to the gradient of the Ricci scalar, expressed as $\nabla_{\nu}T^{\nu}_{\mu}=\lambda R_{,~\mu}$. This formulation leads to modified Einstein field equations (EFE) that are both simple and versatile, offering novel insights into astrophysical and cosmological phenomena. Recent studies have demonstrated the efficacy of Rastall theory in deriving solutions for black holes \cite{Heydarzade}, including charged and rotating NUT black holes \cite{Parihadi}, rotating black holes \cite{Kumar}, Kerr-Newman-AdS black holes \cite{Xu}, black holes influenced by non-commutative geometry \cite{Ma}, and nonlinear charged black holes \cite{Gergess}, as well as other applications. In the context of gravastars, Rastall theory has provided robust and flexible models which contribute to the evaluation of structural and dynamical properties. Ghosh et al. \cite{Bcp} studied the structure and stability of gravastars within the framework of Rastall theory of gravity. Considering the non-conservative nature of Rastall theory, Majeed and Abbas \cite{Majeed} obtained gravastar solutions. Recently, Bhattacharjee and Chattopadhyay \cite{Bhattacharjee2} constructed charged gravastars within Rastall gravity and studied the impact of charge on the gross properties of gravastars. Apart from these, Rastall theory of gravity has been employed in a variety of astrophysical phenomena \cite{Mota,Abbas1,Abbas2,Oliveira,Hanafy,Hansraj,Capone,Batista,Bhar2,Mustafa,Bhattacharjee3}  

Adopting cylindrically symmetric space-time in the modeling of astrophysical compact objects is advantageous for several reasons: (i) it provides a simplified approach to EFE which reduces the computational complexity while retaining the basic physical characteristics, (ii) it helps to understand how the compact objects behave in more complex geometries, (iii) cylindrically symmetric space-times act as an intermediate step between fully spherical and asymmetric space-times. Hence, we can explore the deviations in the properties of compact stellar structures in cylindrical symmetry relative to a spherical or asymmetric space-time. Moreover, cylindrical symmetry has broader implications for understanding the nature of spacetime in modified gravity. It can offer insights into gravitational phenomena and help explore fundamental questions about spacetime structure under modified conservation laws. Now, Rastall theory modifies the non-minimal coupling between matter and geometry, potentially leading to novel solutions. Incorporating cylindrical symmetry in Rastall theory may accent the unique features in ultra-compact objects, such as gravastars. Further, the unique gravity-matter coupling in Rastall theory, expressed through the non-zero divergence of the energy-momentum tensor, may lead to distinct effects which are sensitive to the choice of space-time symmetry. In this regard, cylindrically symmetric space-time provides a less-explored but tractable avenue to test such effects on the gravastar stability, shell mass, and other properties. 

The rest of the paper is organised in the following manner: Section~\ref{sec2} addresses the mathematical formulation of EFE within the context of Rastall gravity, considering a cylindrically symmetric space-time. The main model of gravastar in this formalism is analysed in section~\ref{sec3}. In this section, we have obtained the novel singularity-free and non-vanishing solutions for the three layers of gravastars. Section~\ref{sec4} deals with the modifications of junction conditions from GR due to the incorporation of Rastall theory and cylindrically symmetric space-time. The evaluation of characteristic constants are addressed in section~\ref{sec5} by matching, (i) the interior and shell solutions at the interior radius $(r=r_{1})$ and (ii) the shell and exterior solutions at the outer radius $(r=r_{2})$. In section~\ref{sec6}, we tabulate the mass of the thin shell for different choices of characteristic radii as well as Rastall parameter $(\xi)$. The impact of Rastall parameter $(\xi)$ on the salient properties of gravastars, such as proper length, energy and entropy are discussed in section~\ref{sec7}. The novel approach for the prediction of thin shell mass limit through stability analysis, within this parameter space, is addressed in section~\ref{sec8}. Finally, in section~\ref{sec9}, we summarise the work by discussing the major findings.
  
\section{Rastall theory of gravity and Einstein field equations in cylindrically symmetric space-time}\label{sec2} The novel hypothesis introduced by P. Rastall \cite{Rastall,Rastall1} challenged the idea of the conservation of energy-momentum tensor $(T^{\mu\nu})$ described in Einstein gravity. The preparatory point that $\nabla_{\nu}T^{\mu\nu}\neq0$ allows the energy-momentum tensor of the matter sector to interact with the curvature of space-time in a new way. In this formalism, the non-zero divergence of the energy-momentum tensor is expressed as:
\begin{equation}
	\nabla_{\nu}T^{\nu}_{\mu}=\lambda R_{,\mu}, \label{eq1}
\end{equation}
where, $\lambda$ is a constant, $R$ is the Ricci scalar and $\lambda$ demonstrates the disparity of GR from Rastall theory of gravity. In this framework, the intricate non-minimalistic interplay between the geometry and matter field leads to the following modification of the Einstein tensor and is written as: 
\begin{equation}
	G_{\mu\nu}+k\lambda g_{\mu\nu}R=kT_{\mu\nu}. \label{eq2}
\end{equation}
Within the framework of Rastall gravity, a key aspect is the altered relationship between the energy-momentum tensor $(T_{\mu\nu})$ and the Ricci scalar $(R)$. This connection deviates from the standard form of GR and is expressed as:
\begin{equation}
	R_{\mu\nu}+\Bigg(k\lambda-\frac{1}{2}\Bigg)g_{\mu\nu}R=kT_{\mu\nu}, \label{eq3}
\end{equation}
where, $k$ is the gravitational coupling constant and for simplification, we consider, $k\lambda=\xi$, where, $\xi$ is the dimensionless Rastall parameter.  Now, contraction of Eq.~(\ref{eq3}) leads to $R(4\xi-1)=kT_{\mu}^{\mu}$. Interestingly, for, $\xi=\frac{1}{4}$, we retain the vacuum solutions of EFE with $R_{\mu\nu}=0$ and consequently, $T_{\mu\nu}=0$. In this theoretical framework, using the relativistic units, $G=c=1$, $k$ and $\lambda$ are expressed as:
\begin{equation}
	k=8\pi\Bigg(\frac{4\xi-1}{6\xi-1}\Bigg), \label{eq4}
\end{equation}
and 
\begin{equation}
	\lambda=\frac{\xi(6\xi-1)}{8\pi(4\xi-1)}. \label{eq5}
\end{equation}
Interestingly, setting the coupling constant $\xi=0$ in Eq.~(\ref{eq4}) recovers the Einstein's gravitational constant. However, Eqs.~(\ref{eq4}) and (\ref{eq5}) together predict that $k$ and $\lambda$ diverge when $\xi=\frac{1}{6}$ and $\xi=\frac{1}{4}$ respectively. Consequently, for mathematical consistency, we exclude $\xi=\frac{1}{4}$ and $\xi=\frac{1}{6}$ from further analysis in the present scenario. Given the aforementioned mathematical framework, the Einstein field equations (EFE) are reformulated as:
\begin{equation}
	G_{\mu\nu}+\xi g_{\mu\nu}R=8\pi T_{\mu\nu}\Bigg(\frac{4\xi-1}{6\xi-1}\Bigg). \label{eq6}
\end{equation}
We adopt the most general static, cylindrically symmetric metric introduced by Bronnikov et al. \cite{Bronnikov}, expressed as follows:
\begin{equation}
	ds^{2}=-e^{2\gamma(r)}dt^{2}+e^{2\alpha(r)}dr^{2}+e^{2\zeta}dz^{2}+e^{2\beta}d\phi^{2}, \label{eq7}
\end{equation}
Utilising tangential gauge treatment \cite{Bhattacharjee}, we write $e^{2\beta}=r^{2}$ in Eq.~(\ref{eq7}). Hence, the line element described in Eq.~(\ref{eq7}) becomes dimensionally acceptable and is written as:
\begin{equation}
		ds^{2}=-e^{2\gamma(r)}dt^{2}+e^{2\alpha(r)}dr^{2}+e^{2\zeta}dz^{2}+r^{2}d\phi^{2}, \label{eq8}
\end{equation} 
Since, the coefficients of the $g_{rr}$ and $g_{zz}$ are dimensionally equivalent, we can conveniently consider, $e^{2\alpha(r)}=e^{2\zeta}$ \cite{Bhattacharjee}. Hence, Eq.~(\ref{eq7}) takes the form \cite{Abbas}:
\begin{equation}
		ds^{2}=-e^{2\gamma(r)}dt^{2}+e^{2\alpha(r)}(dr^{2}+dz^{2})+r^{2}d\phi^{2}. \label{eq9}
\end{equation}
The energy-momentum tensor for a perfect fluid interior having isotropic matter distribution is expressed as:
\begin{equation}
	T_{\nu}^{\mu}=diag(-\rho,p,p,p), \label{eq10}
\end{equation}
where, $\rho$ and $p$ are the energy density and pressure respectively. Using Eqs.~(\ref{eq9}) and (\ref{eq10}) in Eq.~(\ref{eq6}), we obtain the following set of tractable equations: 
\begin{eqnarray}
	-\alpha''e^{-2\alpha}+\frac{2\xi e^{-2\alpha}}{r}\Big[\gamma'+r\gamma'^{2}+r(\alpha''+\gamma'')\Big]=8\pi\rho\Bigg(\frac{4\xi-1}{6\xi-1}\Bigg), \label{eq11} \\
	e^{-2\alpha}\Bigg[\alpha'\gamma'+\frac{\alpha'}{r}+\frac{\gamma'}{r}\Bigg]-\frac{2\xi e^{-2\alpha}}{r}\Big[\gamma'+r\gamma'^{2}+r(\alpha''+\gamma'')\Big]=8\pi p\Bigg(\frac{4\xi-1}{6\xi-1}\Bigg), \label{eq12}\\
	e^{-2\alpha}\Bigg[\gamma''+\gamma'^{2}+\frac{\gamma'}{r}-\frac{\alpha'}{r}-\alpha'\gamma'\Bigg]-\frac{2\xi e^{-2\alpha}}{r}\Big[\gamma'+r\gamma'^{2}+r(\alpha''+\gamma'')\Big]=8\pi p\Bigg(\frac{4\xi-1}{6\xi-1}\Bigg), \label{eq13}\\
	e^{-2\alpha}\Bigg[\alpha''+\gamma''+\gamma'^{2}\Bigg]-\frac{2\xi e^{-2\alpha}}{r}\Big[\gamma'+r\gamma'^{2}+r(\alpha''+\gamma'')\Big]=8\pi p\Bigg(\frac{4\xi-1}{6\xi-1}\Bigg), \label{eq14}
\end{eqnarray}
where, the prime $(')$ indicates the derivatives with respect to $r$. Now, ensuring that the Bianchi identity holds in this framework, i.e., $\nabla_{\nu}G^{\mu\nu}=0$, leads to
\begin{equation}
	\frac{p'}{r}+(\rho+p)\gamma'-\frac{\xi}{4\xi-1}(3p'-\rho')=0. \label{eq15}
\end{equation} 
Notably, for $\xi\rightarrow0$, we retain the Einstein theory of gravity. To solve the closed system of Eqs.~(\ref{eq11}), (\ref{eq12}), (\ref{eq13}) and (\ref{eq14}), we use the three distinct EoS for the gravastar system as prescribed below:
\begin{itemize}
	\item Interior region (Region-I): $0\leq r<r_{1}$, $p=-\rho$
	\item Thin shell (Region-II): $r_{1}<r<r_{2}$, $p=\rho$
	\item Exterior region (Region-III): $r>r_{2}$, $p=\rho=0$
\end{itemize}
\begin{figure}
\centering
\includegraphics[width=0.4\textwidth]{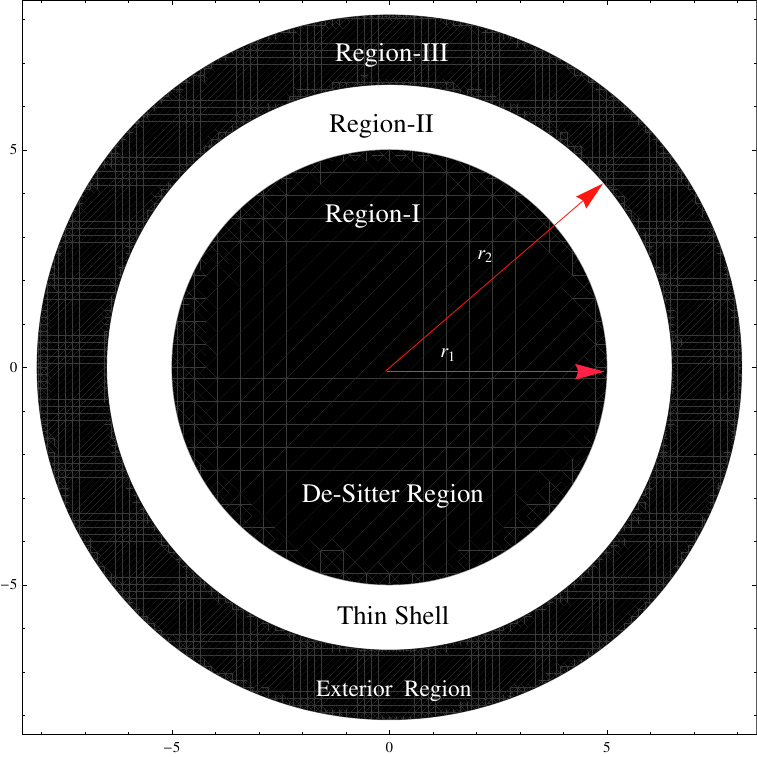}
\caption{Schematic diagram of gravastar}
\label{fig1}
\end{figure}
\section{Gravastar model in Rastall gravity} \label{sec3} We analyse the structure of gravastars in the framework of Rastall gravity in generalised cylindrical symmetry.
\subsection{Interior region} Within the theorised gravastar framework, the interior region is postulated to be akin to a vacuum de-Sitter spacetime. This spacetime is characterised by the EoS expressed as $p=-\rho=\rho_{c}$, where $p$ represents pressure and $\rho$ signifies energy density. This negative pressure engenders a radially outward directed force that acts on all points of the shell structure. This outward force counteracts the inward pull of gravity, thereby preventing gravitational collapse. It is noteworthy that the negative energy density in this region equates to a positive cosmological constant. This specific EoS selection is also referred to as the `degenerate vacuum' or `$\rho$-vacuum' and serves as a prime example of an EoS for dark energy. A linear combination of Eqs.~(\ref{eq11}), (\ref{eq12}), (\ref{eq13}) and (\ref{eq14}) coupled with the EoS results in the following analytical solution:
\begin{eqnarray}
	\gamma=\ln\Big[{c_{1}(r-c_{2})}\Big], \label{eq16} \\
	\alpha=c_{3}, \label{eq17}
\end{eqnarray}
where, $c_{1},~c_{2}$ and $c_{3}$ are non-zero integration constants. Moreover, $c_{1}$ and $c_{2}$ have the dimensions of $Km^{-1}$ and $Km$ respectively while $c_{3}$ is dimensionless. For a physically acceptable solution, we consider the numerical value of $c_{1}=-c_{2}$ in Eq.~(\ref{eq16}). Now, Eqs.~(\ref{eq16}) and (\ref{eq17}) reveal that both the metric components, {\it viz.}, $g_{rr}=e^{2\alpha}$ and $g_{tt}=e^{2\nu}$ remain finite and constant at the centre $(r\rightarrow0)$. This observation contradicts the prediction of a central singularity in the classical black hole model. Consequently, the current framework suggests a possible absence of a singularity at the centre $(r=0)$. The active gravitational mass contained within the interior de-Sitter space-time of radius $r=r_{1}$ is expressed as:
\begin{equation}
	M_{active}=\int_{0}^{r_{1}}4\pi r^{2}\rho dr=\frac{4}{3}\pi r^{3}\rho_{c}, \label{eq18}
\end{equation}    
where, $\rho_{c}$ is the constant interior density. 
\subsection{Thin shell} In gravastars, the intermediate thin shell separates the interior and exterior space-times. The intermediate thin shell contains an ultra-relativistic fluid that adheres to the EoS, $p=\rho$. Zel'dovich \cite{Zeldovich,Zeldovich1} introduced this concept, termed `stiff fluid', within the context of the cold baryonic universe. In this study, we propose that such an approximation can be justified due to either,
\begin{itemize}
	\item thermal fluctuations with zero chemical potential, or 
	\item the conservation of the gravitational quanta number density at zero temperature. 
\end{itemize}
Several researchers \cite{Madsen,Carr,Chakraborty,Buchert,Linares,Braje} have utilised this stiff fluid model to explore different cosmological and astrophysical phenomena. The combination of the non-vacuum thin shell approximation and the stiff fluid equation of state complicates the derivation of exact solutions for this region. Nevertheless, analytical solutions are achievable using the thin shell approximation, i.e., $0<e^{-2\alpha}\ll1$. Compiling the thin shell approximation, thin shell EoS and Eqs.~(\ref{eq11}), (\ref{eq12}), (\ref{eq13}) and (\ref{eq14}), we obtain:
\begin{eqnarray}
	-e^{-2\alpha}\alpha''+2\xi e^{-2\alpha}\alpha''=8\pi\rho\Bigg(\frac{4\xi-1}{6\xi-1}\Bigg), \label{eq19} \\
	e^{-2\alpha}\Bigg[\frac{\alpha'}{r}+\alpha'\gamma'\Bigg]-2\xi e^{-2\alpha}\alpha''=8\pi p\Bigg(\frac{4\xi-1}{6\xi-1}\Bigg), \label{eq20} \\
	e^{-2\alpha}\Bigg[-\frac{\alpha'}{r}-\alpha'\gamma'\Bigg]-2\xi e^{-2\alpha}\alpha''=8\pi p\Bigg(\frac{4\xi-1}{6\xi-1}\Bigg), \label{eq21} \\
	e^{-2\alpha}\alpha''-2\xi e^{-2\alpha}\alpha''=8\pi p\Bigg(\frac{4\xi-1}{6\xi-1}\Bigg). \label{eq22}
\end{eqnarray}
Solving Eqs.~(\ref{eq19}), (\ref{eq20}), (\ref{eq21}) and (\ref{eq22}) by utilising the thin shell EoS, we obtain:
\begin{eqnarray}
	\gamma=\ln\Bigg[{\frac{c_{4}}{r}}\Bigg], \label{eq23} \\
	\alpha=rc_{5}. \label{eq24}
\end{eqnarray} 
Here, $c_{4}$ and $c_{5}$ are non-vanishing integration constants having the dimensions $Km$ and $Km^{-1}$ respectively.  The presence of non-zero curvature invariants $\Big(g_{rr}=e^{2\alpha}$ and $g_{tt}=e^{2\gamma}\Big)$ within this thin shell approximation suggests the absence of an event horizon in the current model. This implies the potential for gravastar configurations as an alternative description to black holes, lacking a region of future light cones entirely trapped within the spacetime. 

Using Eq.~(\ref{eq23}) in Eq.~(\ref{eq15}), the matter density and pressure of the fluid contained within the thin shell is described as:
\begin{equation}
	p=\rho=\rho_{0}\Big[1+2\xi(r-2)\Big]^{\frac{2-8\xi}{2\xi}}. \label{eq25}
\end{equation}
\begin{figure}[h]
	\centering
	\includegraphics[width=0.5\textwidth]{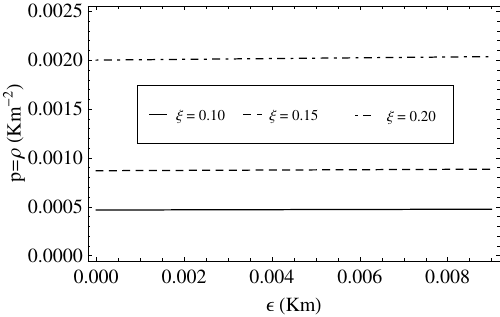}
	\caption{Radial variation of pressure or matter density profile with shell thickness $(\epsilon)$.}
	\label{fig2}
\end{figure}
To describe the impact of Rastall parameter $(\xi)$ on the matter density and pressure within the thin shell, we have plotted the density and pressure profile with shell thickness $(\epsilon)$ in Fig.~\ref{fig2}. From Fig.~\ref{fig2}, we note that for a particular $\xi$, the density increases with increasing shell thickness $(\epsilon)$, i.e., more dense fluid is expected to be found towards the outer edge of the shell. Further, with increasing $\xi$, the gravity-matter coupling significantly increases, as a result the density and pressure in the shell regions amplifies.
\subsection{Exterior region:} The exterior static, vacuum region is characterised by the EoS, $p=\rho=0$. Now, a vacuum solution signifies a space-time devoid of matter and non-gravitational fields. Mathematically, this translates to the vanishing of the Ricci tensor $(R_{\mu\nu}=0)$ throughout the space-time. In this formalism, we obtain the line element for the exterior space-time as \cite{Bhattacharjee}:
\begin{equation}
	ds^{2}=-\Bigg(\frac{c_{7}}{1+rc_{6}}\Bigg)^{2}dt^{2}+e^{2rc_{6}}(dr^{2}+dz^{2}) +r^{2} d\phi^{2}, \label{eq26}
\end{equation}
where, $g_{tt}=e^{2\gamma}=\Big(\frac{c_{7}}{1+rc_{6}}\Big)^{2}$ and $g_{rr}=e^{2\alpha}=e^{2rc_{6}}$. In Eq.~(\ref{eq26}), $c_{6}$ and $c_{7}$ are the integration constants where $c_{6}$ has the dimension of $Km^{-1}$ while $c_{7}$ is dimensionless. Now, following Bhattacharjee et al. \cite{Bhattacharjee}, we have calculated the Kretschmann scalar to  validate the flatness of the exterior geometry. The general form of Kretschmann scalar is expressed as \cite{Henry}:
\begin{equation}
	R_{k}=R^{ABCD}R_{ABCD}, \label{eq27}
\end{equation}
where, $R_{ABCD}$ represent the Riemann curvature tensor. Using Eq.~(\ref{eq26}), we obtain:
\begin{equation}
	R_{k}=\frac{4c_{6}^{2}e^{-4c_{6}r}(4c_{6}^{4}r^{4}+16c_{6}^{3}r^{3}+23c_{6}^{2}r^{2}+10c_{6}r+3)}{r^{2}(1+rc_{6})^{4}}. \label{eq28}
\end{equation} 
It is evident from Eq.~(\ref{eq28}) that as $r\rightarrow\infty$, $R_{k}\rightarrow0$, indicating that the exterior space-time is flat vacuum region. 
\section{Junction condition}\label{sec4} The structure of a charged gravastar is composed of three distinct layers: the interior region, an intermediate thin shell, and the exterior region. The thin shell acts as a boundary separating the interior and exterior geometries. According to the fundamental junction condition, the smooth matching of the interior and exterior metrics at the hyper-surface $(r=R_{b})$ is ensured by the Israel criterion. It is important to note that while the metric coefficients are continuous at this interface, their derivatives may not be. To compute the surface stresses at the junction interface, we employ the Darmois-Israel junction conditions \cite{Israel,Israel1,Darmois}. In the context of Einstein gravity, the Lanczos equation \cite{Lanczos,Sen,Perry,Musgrave}, in coordinates $X^{\kappa} = (t,r,z,\phi)$, is used to express the intrinsic surface energy tensor $(S_{\chi\eta})$ as:
\begin{equation}
	S^{\chi}_{\eta}=-\frac{1}{8\pi}\Bigg[K^{\chi}_{\eta}-\delta^{\chi}_{\eta}K^{\lambda}_{\lambda}\Bigg], \label{eq29}
\end{equation}
where, $K_{\chi\eta}=K_{\chi\eta}^{+}-K_{\chi\eta}^{-}$ is extrinsic curvature tensor describing the exterior $(+)$ and interior $(-)$ interfaces respectively. In the context of modified theories of gravity, the well-known matching conditions employed in GR require adjustments. This work adopts the framework presented in Ref.~\cite{Rosa} to describe the junction condition within Rastall gravity. Notably, the description leverages the distribution formalism to express the Ricci tensor, Ricci scalar, and energy-momentum tensor as:
\begin{eqnarray}
	R_{\mu\nu}=R_{\mu\nu}^{\pm}-\delta(\ell)(\epsilon_{n}[K_{\chi\eta}]e^{\chi}_{\mu}e^{\eta}_{\nu}+n_{\mu}n_{\nu}[K]), \label{eq30} \\
	R=R^{\pm}-\delta(\ell)2\epsilon_{n}[K], \label{eq31} \\
	T_{\mu\nu}=T_{\mu\nu}^{\pm}+\delta(\ell)S_{\mu\nu}, \label{eq32}
\end{eqnarray}
where, $e^{i}_{j}$ represents the projection of a (3+1)-dimensional space-time into a 3-dimensional space-time and $n_{a}n^{a}=\epsilon_{n}$ is the normalisation condition. Additionally, the normalisation constant, $\epsilon_{n}$ is set to $+1$ and $-1$ respectively for time-like and space-like parametrisation. Now, the projection of Eq.~(\ref{eq3}) in 3-dimensions leads to the modification of Lanczos equation, given in Eq.~(\ref{eq29}), in the framework of Rastall theory of gravity and is represented as:
\begin{equation}
	{S}^{\chi}_{\eta}=-\frac{\epsilon_{n}}{8\pi} \Big[{K}^{\chi}_{\eta}+(2\xi-1){\delta}^{\chi}_{\eta}{K}^{\kappa}_{\kappa}\Big], \label{eq33}
\end{equation}  
Now, the interior metric of gravastars have a positive-definite time component which corresponds to a time-like geodesic. This indicates that the time-like paths interior to the gravastar are similar in nature to those in normal space-time configurations. Hence, a gravastar maintains a time-like interior structure which reflects its stability against the gravitational collapse and withstands the potential central singularity arising in the context of a black hole. Therefore, in case of gravastars $\epsilon_{n}=1$ and Eq.~(\ref{eq33}) reduces to 
\begin{equation}
	{S}^{\chi}_{\eta}=-\frac{1}{8\pi} \Big[{K}^{\chi}_{\eta}+(2\xi-1){\delta}^{\chi}_{\eta}{K}^{\kappa}_{\kappa}\Big], \label{eq34}
\end{equation}
where, $\xi=0$ preserves the fundamental structure of Lanczos equations in Einstein gravity and $K_{\chi\eta}=K^{+}_{\chi\eta}-K^{-}_{\chi\eta}$, where the signs $(-)$ and $(+)$ refer to the value of the extrinsic curvature tensor $(K_{\chi\eta})$ at the interior and exterior surfaces, respectively. Now, the expression for the second fundamental $K_{\chi\eta}^{\pm}$ is written as:
\begin{equation}
	K_{\chi\eta}=-{\eta}_{\tau}^{\pm} \Bigg(\frac{\partial^2 {X_{\tau}}}{\partial{\zeta^{\chi}} \partial{\zeta^{\eta}}}+{\Gamma}_{\alpha\beta}^{\tau} \frac{\partial{X_{\alpha}}}{\partial {\zeta^{\chi}}} \frac{\partial{X_{\beta}}}{\partial{\zeta^{\eta}}}\Bigg). \label{eq35}
\end{equation}   
The expression for the double sided normal drawn on the surface is expressed as:
\begin{equation}
	\eta_{\tau}^{\pm}={\pm} \Big|{g}^{\alpha\beta} \frac{\partial{f}}{\partial{x^{\alpha}}} \frac{\partial{f}}{\partial{x^{\beta}}}\Big|^{-\frac{1}{2}}\frac{\partial{f(r)}}{\partial{x^{\tau}}}, \label{eq36}
\end{equation}
with $n^{\tau}n_{\tau}=1$. According to Lanczos equation \cite{Lanczos,Sen,Perry,Musgrave}, the surface energy-momentum tensor at the interface boundary is defined as, ${S}_{\chi\eta}=diag(-\Sigma, \mathfrak{P}, \mathfrak{P}, \mathfrak{P} )$, where $\Sigma$ is the energy density and $\mathfrak{P}$ is the pressure, respectively. Now, in the Framework of Rastall theory of gravity, the expressions for $\Sigma$ and $\mathfrak{P}$ are obtained in the form given below:
\begin{equation}
	\Sigma=\frac{(2\xi-1)}{4\pi R}{\Big(\sqrt{f(r)}\Big)}^{+}_{-}~, \label{eq37} 
\end{equation}
where, $f(r)^{+}_{-}$ represent the $g_{rr}$ metric component of the exterior and interior interfaces respectively. Using Eqs.~(\ref{eq17}) and (\ref{eq26}), the surface energy density is obtained as:
\begin{equation}
	\Sigma=\frac{(2\xi-1)}{4\pi R}\Bigg(e^{c_{6}R}-e^{c_{3}}\Bigg). \label{eq39}
\end{equation}
From Eq.~(\ref{eq39}), it is evident that for $\xi=0$, we obtain the expression of the surface energy-density for a gravastar in a generalised cylindrically symmetric space-time in the framework of Einstein gravity \cite{Bhattacharjee}. Now, the mass of the thin shell is expressed as:
\begin{equation}
	M_{shell}=4\pi R^{2}\Sigma=R(2\xi-1)\Bigg(e^{c_{6}R}-e^{c_{3}}\Bigg). \label{eq40}
\end{equation} 
\section{Boundary Condition}\label{sec5} The determination of the characteristic constants appearing in this formalism is necessary to describe the physical behaviour of the present model. In this segment, we evaluate these constants by, (i) matching the interior and thin shell solutions at $r=r_{1}$ and (ii) matching the thin shell and the exterior solutions at $r=r_{2}$. Additionally, the condition that $M_{shell}>0$ presents a restriction that, for $R>0$, if $\xi>\frac{1}{2}$, $c_{3}<c_{6}R$ and if $\xi<\frac{1}{2}$, $c_{3}>c_{6}R$. Again, from Eq.~(\ref{eq40}), we note that $\xi=\frac{1}{2}$ is not permissible. Following this constraint, we have considered the second set of restrictions, i.e., $\xi<\frac{1}{2}$ and $c_{3}>c_{6}R$. Now, the matching condition reads
\begin{eqnarray}
	e^{2\alpha}=e^{r_{1}c_{5}}, \label{eq41} \\
	c_{1}^{2}(r_{1}-c_{2})^2=\frac{c_{4}^{2}}{r_{1}^{2}}, \label{eq42}
\end{eqnarray}
and,
\begin{eqnarray}
	e^{r_{2}c_{5}}=e^{2r_{2}c_{6}}, \label{eq43} \\
	\frac{c_{4}^{2}}{r_{2}^{2}}=\Big(\frac{c_{7}}{1+rc_{6}}\Big)^{2} \label{eq44}.
\end{eqnarray}
Following Ref.~\cite{Bhattacharjee}, we consider $r_{1}=10$~Km and accounting for the finite slice infinitesimal thickness of the shell, we consider $r_{2}=10.001-10.009$~Km. Considering the characteristic constraints presented in Rastall theory along with those obtained in this approach, we take $\xi=0,~0.2$ and $0.4$. Notably, without any loss of generality, we have assigned the numerical value of 0.00001 to the integration constants $c_{1},~c_{2},c_{6}$ and $c_{7}$ \cite{Bhattacharjee}. We tabulate below the numerical values of the constants $c_{3},~c_{4}$ and $c_{5}$ for three characteristic radii of $9-9.009$ Km, $10-10.009$ Km and $11-11.009$ Km. Notably, we have considered the values of $c_{3}$ based on the chosen restrictions as described earlier.  
\begin{table}[h!]
	\centering
	\caption{Numerical evaluation of the necessary constants.}
	\label{tab1}
\begin{tabular}{cccc}
	\hline
	Radius (Km)& $c_{3}$ & $c_{4}$ & $c_{5}$\\
	\hline
	\multirow{3}{*}{9-9.009}& 0.10 & \multirow{3}{*}{0.00045} & 0.005  \\
	&0.30	& & 0.016 \\
	&0.50 & & 0.027 \\
	\hline
	\multirow{3}{*}{10-10.009}& 0.10 & \multirow{3}{*}{0.00055} & 0.005  \\
	&0.30	& & 0.015 \\
	&0.50 & & 0.025 \\
	\hline
	\multirow{3}{*}{11-11.009}& 0.10 & \multirow{3}{*}{0.00066} & 0.004  \\
	&0.30	& & 0.013 \\
	&0.50 & & 0.022 \\
	\hline
\end{tabular}
\end{table}    
\section{Determination of mass of the thin shell}\label{sec6} In this section, we determine the mass of the thin shell by using Eq.~(\ref{eq40}) for the different choices of Rastall parameter $(\xi)$. Here, we utilise the numerical data provided in Table~\ref{tab1} and the results are summarised in Table~\ref{tab2}.
\begin{table}[h!]
	\centering
	\caption{Determination of the mass of the thin shell.}
	\label{tab2}
	\begin{tabular}{ccccc}
		\hline
		\multirow{2}{*}{$\xi$} & \multirow{2}{*}{$c_{3}$} & $M_{shell} (M_{\odot})$ & $M_{shell} (M_{\odot})$ & $M_{shell} (M_{\odot})$\\
		& & $(R=9.009)$ & $(R=10.009)$ & $(R=11.009)$ \\
		\hline
		\multirow{3}{*}{0} & 0.10 & 0.64 & 0.71 & 0.78 \\
		& 0.30 & 2.13 & 2.37 & 2.61 \\
		& 0.50 & 3.96 & 4.40 & 4.84 \\
		\hline
		\multirow{3}{*}{0.10} & 0.10 & 0.51 & 0.57 & 0.63 \\
		                      & 0.30 & 1.71 & 1.89 & 2.08 \\
		                      & 0.50 & 3.17 & 3.52 & 3.87 \\
		\hline
		\multirow{3}{*}{0.15} & 0.10 & 0.45 & 0.50 & 0.55 \\
							  & 0.30 & 1.49 & 1.66 & 1.83 \\
							  & 0.50 & 2.77 & 3.08 & 3.38 \\
		\hline					
		\multirow{3}{*}{0.20} & 0.10 & 0.38 & 0.43 & 0.47 \\
		& 0.30 & 1.28 & 1.42 & 1.56 \\
		& 0.50 & 2.38 & 2.64 & 2.90 \\
		\hline
	\end{tabular}
\end{table}
From Table~\ref{tab2}, it is evident that the mass of the thin shell $(M_{shell})$ increases with the increasing radius for a fixed values of $c_{3}$. Again, for a fixed value of the characteristic radius, $M_{shell}$ increases with increasing $c_{3}$. Notably, Rastall parameter $(\xi)$ has a prominent effect on $M_{shell}$. Here, $\xi=0$ reproduces the previously obtained results of gravastar formalism in cylindrical symmetry in the framework of Einstein gravity by Bhattacharjee et al. \cite{Bhattacharjee}. On the other hand, as $\xi$ increases, the gravity-matter coupling becomes stronger and the mass of the thin shell decreases which is also evident from Table~\ref{tab2}. 
\section{Key attributes of gravastars in this model}\label{sec7} This section focuses on the quantitative assessment of the key attributes within the gravastar model in a generalised cylindrically symmetric space-time within the framework of Rastall gravity. The analysis will encompass the proper length, energy, entropy and EoS parameter of the thin shell. Additionally, the variation of these properties as a function of shell thickness will be visualised through the generation of graphical representations.
\subsection{Proper length} 
\begin{figure}[h]
	\centering
	\includegraphics[width=0.5\textwidth]{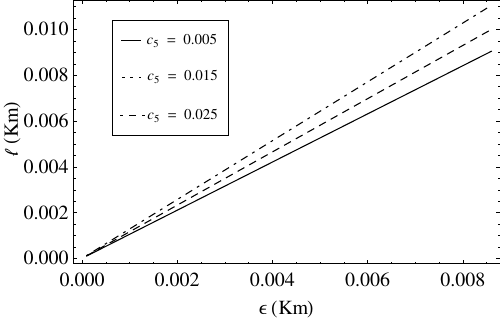}
	\caption{Variation of proper length $(\ell)$ with shell thickness $(\epsilon)$.}
	\label{fig3}
\end{figure}

The thickness of the thin shell is infinitesimally small $(\epsilon=r_{2}-r_{1}\ll1)$ \cite{Mazur,Mazur1,Mazur2}. Using Eq.~(\ref{eq24}), the proper length is expressed as:
\begin{equation}
	\ell=\int_{r_{1}}^{r_{2}}e^{\alpha}dr=\frac{(e^{c_{5}}r_{2}-e^{c_{5}}r_{1})}{c_{5}}. \label{eq45}
\end{equation}
Interestingly, in this formalism, Eq.~(\ref{eq45}) is independent of $\xi$. Therefore, the proper length of the thin shell of gravastars in a generalised cylindrically symmetric space-time within the framework of Rastall gravity is similar to that in Einstein gravity. However, using Table~\ref{tab1}, we have graphically represented the variation of proper length with the shell thickness in Fig.~\ref{fig3}. 
It is noted that, with increasing shell thickness $(\epsilon)$ and the constant $c_{5}$, the proper length $(\ell)$ increases. 
\subsection{Energy} 
\begin{figure}[h]
	\centering
	\includegraphics[width=0.5\textwidth]{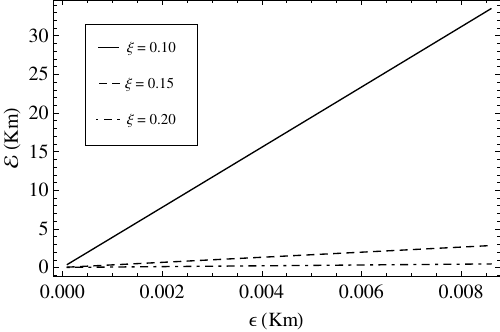}
	\caption{Variation of energy $(\mathcal{E})$ with shell thickness $(\epsilon)$.}
	\label{fig4}
\end{figure}

The Mazur-Mottola model \cite{Mazur,Mazur1,Mazur2} suggests that the shell region is described by an ultra-relativistic stiff fluid having a particular EoS, $p=\rho$, which is a special instance of the barotropic EoS, $p=\omega\rho$, where $\omega=1$. This form of EoS demonstrates the extreme limit of causality and it was first put forward by Zel'dovich \cite{Zeldovich,Zeldovich1} in the context of cold baryonic universe. Now, the energy contained within the thin shell is expressed as:
\begin{equation}
	\mathcal{E}=\int_{r_{1}}^{r_{2}} 4\pi r^2\rho~dr. \label{eq46}
\end{equation}     
Now, using Eq.~(\ref{eq25}), we have integrated Eq.~(\ref{eq46}) to obtain the analytical expression for the energy of the thin shell. However, the variation of shell energy $(\mathcal{E})$ with respect to the shell thickness $(\epsilon)$ for different Rastall parameters is shown Fig.~\ref{fig4}.   
From Fig.~\ref{fig4}, it is evident that more energetic fluid is found towards the outer boundary. Moreover, as $\xi$ increases, the strength of the non-minimal coupling, presented through Rastall theory, increases. Consequently, the energy of the thin shell decreases which is illustrated in Fig.~\ref{fig4}.  
\subsection{Entropy} Entropy quantifies the disorder within mechanical systems. The Mazur-Mottola model \cite{Mazur,Mazur1,Mazur2} posits that the entropy of the vacuum interior is null, and the thin shell is the sole contributor to entropy considerations. The entropy is assessed using an entropy function of the following form:
\begin{equation}
	S=4\pi\int_{r_{1}}^{r_{2}} \mathfrak{s}(r)r^2e^{\alpha} dr, \label{eq47}
\end{equation}
where Here $\mathfrak{s}(r)=\frac{\psi^2k_{B}^2T(r)}{4\pi\hbar^2}=\frac{\psi k_B}{\hbar}\sqrt{\frac{p}{2\pi}}$ represents the entropy density function associated with the local temperature $T(r)$, $\psi$ is a dimensionless function which is taken to be unity without any loss of generality, $k_{B}$ represents the Boltzmann constant and $\hbar=\frac{h}{2\pi}$ is the reduced Planck's constant. Substituting Eqs.~(\ref{eq24}) and (\ref{eq25}) in Eq.~(\ref{eq47}), we obtain the expression for the entropy of the fluid contained within the thin shell. However, to avoid the mathematical complexity, we have opted for the graphical representation of the variation of entropy with respect to shell thickness. 
\begin{figure}[h!]
	\centering
	\begin{subfigure}[t]{0.45\textwidth}
		\centering
		\includegraphics[width=1\textwidth]{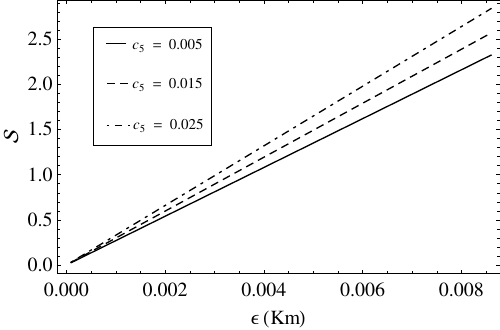}
		\caption{}
	\end{subfigure}
	\hfill
	\begin{subfigure}[t]{0.45\textwidth}
		\centering
		\includegraphics[width=1\textwidth]{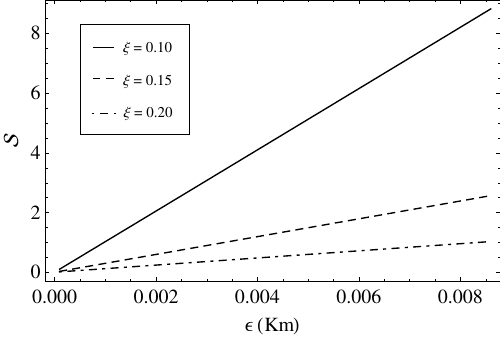}
		\caption{}
	\end{subfigure}
	\caption{Variation of shell entropy $S$ with shell thickness $\epsilon$~(Km) for (a) fixed value of $\zeta=0.15$ and (b) fixed value of $c_{5}=0.015$.\label{fig5}}   
\end{figure}
Fig.~\ref{fig5} illustrates the nature of shell entropy for two scenarios. From the figure on the left panel, it is evident that with increasing $c_{5}$, the entropy increases. While from the right panel, we observe that, as the Rastall parameter increases, the coupling between gravity and matter increases. Hence, comparatively a more compact shell structure can be achieved, resulting in a decrease in the entropy of the thin shell.  
\section{Analysing stability through surface redshift and prediction of mass limit}\label{sec8} The gravitational surface redshift $(Z_{s})$ is a key indicator in assessing the stability of stellar configurations. It is defined as $Z_{s}=\frac{\lambda_{e}-\lambda_{o}}{\lambda_{o}}$, where, $\lambda_{e}$ is the wavelength emitted by the source and $\lambda_{o}$ is the wavelength observed. Now, Buchdahl established that for a perfect isotropic fluid distribution, the surface redshift must satisfy $Z_{s}<2$ \cite{Buchdahl,Straumann}. B\"ohmer and Harko \cite{Bohmer} extended this concept for anisotropic stars in the presence of a cosmological constant, determining that, $Z_{s}\leq5$. Ivanov \cite{Ivanov} further refined the upper limit of surface redshift, placing it at $Z_{s}\leq5.211$. The surface redshift is expressed as:
\begin{equation}
	1+Z_{s}=\Bigg(1-\frac{2M_{shell}}{R}\Bigg)^{-\frac{1}{2}}, \label{eq48}
\end{equation}
where, $M_{shell}$ defines the mass of the thin shell and $R$ is the characteristic radius. Using Eq.~(\ref{eq40}) in Eq.~(\ref{eq48}), we have numerically evaluated the values of surface redshift $(Z_{s})$ for different choices of Rastall parameter $(\xi)$ permissible in this framework and are tabulated in Table~\ref{tab3}. 
\begin{table}[h]
\centering
\caption{Numerical evaluation of surface redshift $(Z_{s})$.}
\label{tab3}
\begin{tabular}{ccccc}
\hline
\multirow{2}{*}{Rastall parameter $(\xi)$} & \multirow{2}{*}{$c_{3}$} & \multicolumn{3}{c}{Surface Redshift $(Z_{s})$} \\ \cline{3-5} 
& & $R=9.009~Km$ & $R=10.009~Km$ & $R=11.009~Km$ \\
\hline
\multirow{4}{*}{0} & 0.10 & 0.12520 & 0.12518 & 0.12517 \\
				   & 0.30 & 0.82433 & 0.82427 & 0.82421 \\
				   & 0.35 & 1.48417 & 1.48402 & 1.48387 \\
				   & 0.40 & 6.77774 & 6.77303 & 6.76834 \\
				   \hline
\multirow{4}{*}{0.10} & 0.10 & 0.09641 & 0.09639	& 0.09638 \\
					  & 0.30 & 0.506923 & 0.506896 & 0.506868 \\
					  & 0.40 & 1.16562 & 1.16553 & 1.16545 \\
					  & 0.45 & 2.3178 & 2.31751 & 2.31722 \\			   
				   \hline
\multirow{3}{*}{0.15} & 0.10 & 0.0828151 & 0.0828062 & 0.0827973\\
					  & 0.30 & 0.399836 & 0.399817 & 0.399797 \\
					  & 0.50 & 2.2984 & 2.2981 & 2.2979 \\ 
					  \hline
\multirow{4}{*}{0.20} & 0.10 & 0.069716 & 0.069708 & 0.069701 \\
					  & 0.30 & 0.31275 & 0.312737 & 0.312723 \\
					  & 0.50 & 1.12409 & 1.12403 & 1.12398 \\
					  & 0.60 & 7.58581 & 7.58201 & 7.57822 \\
 	 			     \hline			   
\end{tabular}
\end{table} 
In Table~\ref{tab3}, the choice of $c_{3}$ is very interesting. We have noted that for these particular choices of Rastall parameter $(\xi)$, $c_{3}$ has an upper limit to ensure that $Z_{s}$ remains real. Exceeding such values of $c_{3}$ gives real mass values but the condition that $\frac{2M_{shell}}{R}<1$ is violated. Hence, we have restricted the values of $c_{3}$ for different Rastall parameters in Table~\ref{tab3}. Now, using Table~\ref{tab3} and the absolute upper limit of Buchdahl surface redshift bound, i.e., $Z_{s}=2$, we obtain an estimation of the mass of the thin shell for different characteristic radii without violating the stability criterion and the results are tabulated in Table~\ref{tab4}. 
\begin{table}[h]
	\centering
	\caption{Estimation of mass of the thin shell $(M_{shell})$ from the analysis of surface redshift.}
	\label{tab4}
\begin{tabular}{ccccc}
	\hline
	\multirow{2}{*}{Rastall parameter $(\xi)$} & \multicolumn{3}{c}{$M_{shell}~(M_{\odot})$} & \multirow{2}{*}{$c_{3}$}\\ \cline{2-4} 
	& $R=9.009~(Km)$ & $R=10.009~(Km)$ & $R=11.009~(Km)$ &  \\
	\hline
	0 & 2.716 & 3.018 & 3.319 & $\leq0.368$ \\
	0.10 & 2.715 & 3.016 & 3.318 & $\leq0.442$ \\
	0.15 & 2.710 & 3.011 & 3.311 & $\leq0.491$ \\
	0.20 & 2.712 & 3.013 & 3.314 & $\leq0.554$ \\
	\hline
\end{tabular}
\end{table} 
Interestingly, from Table~\ref{tab4}, we note that with increasing $\xi$, the non-minimal coupling between matter and gravity increases. Therefore, the mass decreases. However, for a particular $\xi$, the mass increases with increasing radius. Additionally, it is evident that for $\xi=0.20$, the mass of the thin shell increases again. To investigate this increased mass value, we have calculated the compactness $(u=\frac{M_{shell}}{R})$ of the present model and the results are tabulated in Table~\ref{tab5} for different choices of Rastall parameter $(\xi)$ and characteristic radii. 
\begin{table}[h]
	\centering
	\caption{Estimation of mass of the thin shell $(M_{shell})$ from the analysis of compactness.}
	\label{tab5}
	\begin{tabular}{cccc}
		\hline
		\multirow{2}{*}{Rastall parameter $(\xi)$} & \multicolumn{3}{c}{Compactness $\Big(u=\frac{M_{shell}}{R}\Big)$} \\ \cline{2-4}
		& $R=9.009~(Km)$ & $R=10.009~(Km)$ & $R=11.009~(Km)$ \\
		\hline 
		0 & 0.444752 & 0.444742 & 0.444732 \\
		0.10 & 0.444581 & 0.444573 & 0.444565 \\
		0.15 & 0.443701 & 0.443694 & 0.443687 \\
		0.20 & 0.444066 & 0.444060 & 0.444054 \\
		\hline
	\end{tabular}
\end{table}  
It is evident from Table~\ref{tab5} that the compactness increases for $\xi=0.20$. This may be attributed to the fact that, as $\xi$ increases, the pressure and density distribution of the gravastar change which leads to a denser configuration. The dense structure along with the ultra-relativistic stiff fluid contained within the thin shell may point towards an increased compactness. Moreover, since, gravastars are supposed to be the alternative manifestation of the black holes, the compactness limit $(u\sim0.5)$ is also well obeyed in the present framework. From the current analysis, it is clear that Table~\ref{tab4} provides estimates for the mass of the thin shell, corresponding to various choices of the Rastall parameter $(\xi)$ and characteristic radii $(R)$, while ensuring compliance with the stability criterion.

\section{Discussion}\label{sec9}
In this research, we provide an estimation for the mass of the gravastar shell through a novel isotropic gravastar solution within the framework of Rastall gravity, considering a generalised cylindrically symmetric space-time \cite{Bhattacharjee}. A tractable set of EFE, represented in Eqs.~(\ref{eq11}), (\ref{eq12}), (\ref{eq13}), and (\ref{eq14}), has been constructed to derive an exact set of physically consistent solutions for the three distinct layers of the isotropic gravastar configuration. Additionally, we have analysed the fundamental structure of gravastars, finding the results to align closely with the Mazur-Mottola model \cite{Mazur,Mazur1,Mazur2}. We have also investigated the influence of Rastall parameter $(\xi)$ on the structure and dynamics of gravastars. Furthermore, several noteworthy features have been identified and are discussed below:
\begin{itemize}
	\item {\bf Interior region:} The interior region is characterised by an EoS of the form $p=-\rho$, which is referred to as a `$\rho-$vacuum' or `degenerate vacuum' and it implies a dark energy EoS. We have solved Eqs.~(\ref{eq11}), (\ref{eq12}), (\ref{eq13}), and (\ref{eq14}) using this EoS and obtained non-singular solutions for the $g_{rr}~(=e^{2\alpha})$ and $g_{tt}~(=e^{2\gamma})$ metric components on the interior region, as expressed in Eqs.~(\ref{eq16}) and (\ref{eq17}). Hence, the problem of central singularity, persistent in classical black holes, is effectively removed in case of gravastars in this formalism. Furthermore, in Eq.~(\ref{eq18}), we have formulated an analytical expression for the active gravitational mass contained within this region. 
	\item {\bf Thin shell:} The finite slice of thickness of the thin shell separates the interior from the exterior region in a gravastar model. The fluid in this region is defined by the EoS, $p=\rho$, representing an ultra-relativistic stiff fluid in the extreme end of causality. To obtain the analytical solutions, we have approximated the entire set of EFE for the infinitesimally small thickness of the shell and they are expressed in Eqs.~(\ref{eq19}), (\ref{eq20}), (\ref{eq21}) and (\ref{eq22}). Coupling this set of equations with the thin shell EoS, we have obtained the non-vanishing shell solutions as described in Eqs.~(\ref{eq23}) and (\ref{eq24}). This non-vanishing nature demonstrates the absence of event horizon in gravastars relative to black holes. In this framework, the concept of the non-conservation of the energy-momentum tensor, as formulated in Eq.(\ref{eq15}), has been utilised to derive the expressions for matter density and pressure within the shell region, as described by Eq.(\ref{eq25}). To study the impact of Rastall parameter $(\xi)$ on matter density and pressure, we have graphically illustrated the variation of energy density and pressure of the matter contained within the shell region with shell thickness $(\epsilon)$ in Fig.~\ref{fig2}. From Fig.~\ref{fig2}, we note that for a fixed value of $\xi$, the density increases with increasing shell thickness $(\epsilon)$, indicating a higher concentration of matter toward the outer edge of the shell. Additionally, the non-conservation of energy-momentum tensor leads to a redistribution of energy density and pressure in the system. Now, as $\xi$ increases, the coupling between gravity and matter intensifies and this redistribution tends to concentrate energy and pressure more intensely in the shell region. Hence, with increasing $\xi$, energy density and pressure of the fluid contained within the thin shell increase. 
	\item {\bf Exterior region:} The exterior vacuum region is characterised by $p=\rho=0$. Following the work of Bhattacharjee et al. \cite{Bhattacharjee}, we have computed the Kretschmann scalar, expressed in Eq.~(\ref{eq28}), for the exterior region to obtain a vacuum flat space-time. 
	\item {\bf Junction condition:} Within this parameter space, we have revised the junction conditions and found that, for a vanishing Rastall parameter $(\xi\rightarrow0)$, these conditions converge to those derived from Einstein gravity. The interior and exterior spacetime solutions are seamlessly connected at the smooth hypersurface $(r=R)$, adhering to the Israel junction conditions \cite{Israel,Israel1}. Using this matching procedure, we have determined the intrinsic surface energy tensor. The Lanczos formulation \cite{Lanczos,Sen,Perry,Musgrave} was employed to derive the surface energy density, as expressed in Eq.~(\ref{eq37}), which in turn were utilised to define the mass of the thin shell of gravastar, as shown in Eq.~(\ref{eq40}). Again, for $\xi\rightarrow0$, we obtain the expression for the thin shell mass as described in Bhattacharjee et al. \cite{Bhattacharjee} for a generalised cylindrically symmetric space-time. 
	\item {\bf Evaluating constants through boundary condition:} Following Fig.~\ref{fig1}, we evaluate the necessary constants of the model by accounting for the concentric layered structure of a gravastar. Hence, we match the interior and shell solutions at $r=r_{1}$ and at $r=r_{2}$ we match the shell and exterior solutions. The condition $M_{shell}>0$ imposes a constraint on the characteristic constants such that, for $R>0$, if $\xi>\frac{1}{2}$, then $c_{3}<c_{6}R$, whereas if $\xi<\frac{1}{2}$, it follows that $c_{3}>c_{6}R$. Furthermore, from Eq.~(\ref{eq40}), it is evident that the case $\xi=\frac{1}{2}$ is not viable. Taking into account the characteristic constraints derived from Rastall theory, along with those established in this framework, we consider the values $\xi=0.10,~0.15$ and $0.20$. Adhering to these conditions, we have focused on the second set of restrictions. Now, considering the infinitesimally small thickness of the shell, we have chosen three characteristic radii for this model, {\it viz.}, 9-9.009 Km, 10-0.009 Km and 11-11.009 Km \cite{Bhattacharjee}. The numerical values of the constants for these radii are tabulated in Table~\ref{tab1}. 
	\item {\bf Mass of the thin shell:} Using the numerical values of Table~\ref{tab1} in Eq.~(\ref{eq40}), we have tabulated the mass of the thin shell in Table~\ref{tab2}. We note the following key features in this determination, {\it viz.}, (i) for fixed values of $\xi$ and $c_{3}$, $M_{shell}$ increases with increasing radii $(R)$, (ii) for fixed values of $\xi$ and $R$, $M_{shell}$ increases with increasing $c_{3}$, and (iii) finally for fixed values of $R$ and $c_{3}$, we note that with increasing $\xi$, $M_{shell}$ decreases. This may be explained through the fact that with increasing $\xi$, the gravity-matter coupling increases which leads to the redistribution of energy density and pressure. Through this redistribution and increased coupling, the gravitational pull becomes dominant relative to the pressure component. Therefore, the mass decreases. Notably, for $\xi=0$, the results of Table~\ref{tab1} resembles the results obtained in Ref.~\cite{Bhattacharjee}. 
	\item {\bf Key features of Gravastar:} We have studied the basic characteristic features of gravastar in this formalism, such as proper length $(\ell)$, energy $(\mathcal{E})$ and entropy $(S)$ of the thin shell.
	\begin{itemize}
		\item {\bf Proper length of the shell:} In Eq.~(\ref{eq45}), we have obtained the analytical form of the proper length $(\ell)$ and notably, $(\ell)$ is independent of Rastall parameter $(\xi)$. Consequently, the proper length of the thin shell of gravastars in a generalised cylindrically symmetric spacetime within the framework of Rastall gravity aligns with its counterpart in Einstein gravity. Utilising the data from Table~\ref{tab1}, we have illustrated the variation of the proper length with respect to the shell thickness $(\epsilon)$ in Fig.~\ref{fig3}. We note that with increasing shell thickness $(\epsilon)$ the proper length increases. 
		\item {\bf Energy of the shell:} Using Eq.~(\ref{eq25}), we have analytically solved Eq.~(\ref{eq46}) to derive an expression for the energy of the thin shell. The relationship between the shell energy $(\mathcal{E})$ and the shell thickness $(\epsilon)$ for various values of the Rastall parameter $(\xi)$ is depicted in Fig.~\ref{fig4}. From this figure, it is clear that the energy of the fluid increases towards the outer boundary of the shell. Additionally, as $\xi$ grows, representing a stronger non-minimal coupling in the framework of Rastall theory, the energy of the thin shell correspondingly decreases, as demonstrated in Fig.~\ref{fig4}.
		\item {\bf Entropy of the shell:} Fig.~\ref{fig5} presents the behaviour of shell entropy under two distinct scenarios. The left panel demonstrates that entropy increases with higher values of $c_{5}$. Conversely, the right panel reveals that an increase in the Rastall parameter $(\xi)$ enhances the coupling between gravity and matter. As a result, the shell becomes more compact, leading to a reduction in its entropy.
	\end{itemize} 
	\item {\bf Stability and determination of mass limit:} Stability of the present model is analysed on the basis of surface redshift calculation. Buchdahl \cite{Buchdahl} showed that for an isotropic fluid sphere, the surface redshift $(Z_{s})$ must obey the limit, $Z_{s}\leq2$. Following this notion, we have calculated $Z_{s}$ within the parameter space and the results are tabulated in Table~\ref{tab3}. Using Eq.~(\ref{eq40}) in Eq.~(\ref{eq48}), we numerically computed the surface redshift $(Z_{s})$ for various Rastall parameter $(\xi)$ values within the allowed framework, as shown in Table~\ref{tab3}. Notably, the parameter $c_{3}$ exhibits an upper limit to ensure that $Z_{s}$ remains real. If $c_{3}$ exceeds this limit, the condition $\frac{2M_{shell}}{R} < 1$ is violated, despite producing real mass values. Consequently, the permissible $c_{3}$ values for different $\xi$ are restricted in Table~\ref{tab3}. Using the maximum Buchdahl surface redshift $(Z_{s}=2)$, we estimated the thin shell's mass for varying characteristic radii while satisfying the stability criterion, with the results detailed in Table~\ref{tab4}. Notably, for $\xi=0$, the upper limit of surface redshift, $Z_{s}=2$, corresponds to the maximum allowed value of $c_{3}$ which is 0.368 and accordingly, the mass of the thin shell for such configuration is $2.716,~3.018$ and $3.319~M_{\odot}$ respectively for characteristic radii of $9.009,~10.009$ and $11.009~Km$ respectively. Further, it is observed that as $\xi$ increases, the non-minimal coupling between gravity and matter strengthens, leading to a decrease in mass. However, for a fixed $\xi$, the mass grows with increasing radius. Interestingly, for $\xi=0.20$, the shell mass increases again. To understand this, we analysed the compactness $(u=\frac{M_{shell}}{R})$ of the model, as presented in Table~\ref{tab5}. The results indicate that compactness increases for $\xi=0.20$, attributed to changes in the pressure and density profiles, resulting in a denser configuration. This dense structure, combined with the ultra-relativistic stiff fluid in the shell, explains the increased compactness. Furthermore, the compactness remains within the theoretical limit $(u\sim0.5)$, consistent with gravastars being viable black hole alternatives. Thus, Table~\ref{tab4} provides reliable mass estimates for the thin shell across different $\xi$ and radii, adhering to stability conditions.   
\end{itemize}

\section{Acknowledgments} DB is thankful to the Department of Science and Technology (DST), Govt. of India, for providing the fellowship vide no: DST/INSPIRE Fellowship/2021/IF210761. PKC gratefully acknowledges support from the Inter-University Centre for Astronomy and Astrophysics (IUCAA), Pune, India, under Visiting Associateship programme. 

\bibliographystyle{elsarticle-num} 


\end{document}